\documentstyle[psfig]{lamuphys}
\makeatletter
\let\chapter\hid@chapter
\makeatother

\begin{document}
\pagenumbering{arabic}
\title{Enrichment of the Intracluster Medium}

\author{Daniel Thomas}

\institute{Universit\"ats-Sternwarte M\"unchen, Scheinerstra{\ss}e 1,
D-81679 M\"unchen, Germany}

\maketitle

\begin{abstract}
The relevance of galaxies of different luminosity and mass for the chemical
enrichment of the intracluster medium (ICM) is analysed. For this purpose, I
adopt the composite luminosity function of cluster galaxies from Trentham
(1998), which exhibits a significant rise at the very faint end. The model
-- adopting a universal Salpeter IMF -- is calibrated on reproducing the
$M_{\rm\-Fe}/L_{\rm\-tot}$, $M_{\rm\-Fe}/M_*$, and $\alpha$/Fe ratios
observed in clusters. Although the contribution to total luminosity and ICM
metals peaks around $L^*$ galaxies ($M^*\approx -20$), faint objects with
$M_B>-18$ still provide at least 30 per cent of the metals present in the
ICM. In consistency with the solar $\alpha$/Fe ratios determined by {\em
ASCA}, the model predicts that $\sim 60$ per cent of the ICM iron comes from
Type~Ia supernovae. The predicted slope of the relation between intracluster
gas mass and cluster luminosity emerges shallower than the observed one,
indicating that the fraction of primordial gas increases with cluster
richness.
\end{abstract}

\section{Introduction}
In their comprehensive analysis of cluster properties, \cite{Aetal92} find
that the total mass of iron ($M_{\rm\-Fe}^{\rm\-ICM}$) in the intracluster
medium (ICM) is directly proportional to the luminosity of E/S0 cluster
members. In particular there is no correlation with the spiral population.
The logical conclusion is that ellipticals and lenticular galaxies dominate
the chemical enrichment of the ICM.

On the other hand, there are several indications for additional ICM sources.
At the very faint end, the galaxy luminosity function (LF) does not follow
the simple \cite{Sch76} shape, but steepens for objects fainter than
$M_B\approx\--14$ (e.g.\ Driver et al.\ 1994)\nocite{Detal94}. \cite{Tr94}
even concludes that the entire ICM gas may originate from dwarf systems,
which is in turn doubted by \cite{GM97}. The role of dwarf galaxies for the
ICM enrichment is thus still a controversial issue. Moreover, intergalactic
stars which are stripped from galaxies theoretically are predicted to
contain 10--70 per cent of the cluster mass (Moore, Katz, and Lake
1996)\nocite{MKL96}. Indeed, intergalactic planetary nebulae (Mendez et al.\
1997)\nocite{Metal97} and RGB stars (Ferguson et al.\ 1998)\nocite{FTH98}
have been found in the Virgo cluster. Finally, semi-analytic models of
hierarchical galaxy formation predict that dwarf galaxies deliver up to 40
per cent of the ICM metals (Kauffmann \& Charlot 1998)\nocite{KC98}.

The aim of this paper is to analyse the relative contributions from galaxies
of different luminosity and mass to the ICM enrichment for a given LF, which
is adopted from \cite{Tr98}. The LF can be fit by the following double-power
law ($M_B^*=-20.3$, $M_{\rm t}=-14.1$, $\alpha=-1.2$,
$\beta=-0.7^{+0.4}_{-0.2}$):
\begin{equation}
\phi(L)\sim\: (L/L^*)^{\alpha}\, \exp(-L/L^*)\; [1+(L/L_{\rm t})^{\beta}]
\label{eq:phi}
\end{equation}
The slope $\alpha+\beta=-1.9$ at the very faint end causes a steep increase
of the {\em number} of dwarf galaxies, and agrees well with other
determinations in the literature (i.e.\ Zucca et al.\ 1997)\nocite{Zetal97}.
The code of chemical evolution used for the analysis is calibrated on
abundance features in the solar neighbourhood and in elliptical galaxies
(Thomas, Greggio, and Bender 1998a,b)\nocite{TGB98a,TGB98b}. The slope
above $1~M_{\odot}$ of the IMF is assumed universally Salpeter ($x=1.35$).

\section{Galaxy-cluster asymmetry}
A problem that arises if giant ellipticals (gE) alone enrich the ICM is the
so-called galaxy-cluster asymmetry, which predicts sub-solar $\alpha$/Fe
ratios in the ICM in case of $\alpha$-enhanced stellar populations in gEs
(Renzini et al. 1993)\nocite{Retal93}.
\begin{figure}[ht]
\psfig{figure=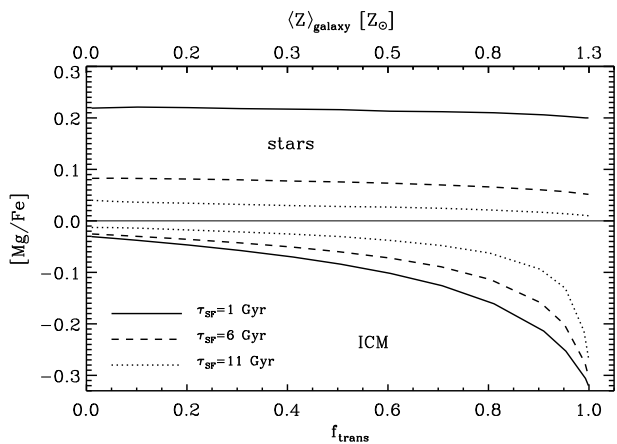,width=0.7\linewidth}
\caption{Mg/Fe ratios in the stars (above the thin line) and the ISM (below
the thin line) as function of transformed gas fraction (lower x-axis) and
respective stellar metallicity (upper x-axis). The results for three
different star formation time-scales $\tau_{\rm SF}$ of galaxy evolution are
shown.}
\label{fig:asym}
\end{figure}

This feature is shown in Fig.~\ref{fig:asym} where I plot the Mg/Fe ratios
of the stars and of the interstellar medium (ISM) as a function of different
galaxy metallicities (upper x-axis), hence mass or luminosity. The larger
the gas fraction $f_{\rm trans}$ (lower x-axis) that is transformed into
stars in a galaxy is, the higher becomes its metallicity. The remaining
interstellar gas plus the re-ejected metal-enriched fraction from the stars
(including SNIa) establishes the ISM, which is assumed to be blown out of
the galaxy in a galactic wind. Galaxies that host Mg/Fe {\em overabundant}
stellar populations owing to their short star formation time-scales
$\tau_{\rm SF}$, in turn enrich the ICM with {\em underabundant} Mg/Fe
ratios. This pattern is particularly prominent for the most massive
(metal-rich) gEs, and stands in conflict to recent data from {\em ASCA} that
constrain roughly solar $\alpha$/Fe ratios in the ICM (Mushotzky et al.\
1996; Ishimaru \& Arimoto 1997)\nocite{Metal96,IA97}. A potential solution
to this inconsistency is the assumption of a top-heavy IMF in gEs, assigning
the main role of ICM enrichment (including Fe) to SNII (Gibson, Loewenstein
\& Mushotzky 1997)\nocite{GLM97}. Fig.~\ref{fig:asym} shows that
alternatively, metal-poor galaxies (i.e.\ dwarf galaxies) eject material
that exhibits roughly solar Mg/Fe ratios, independent of their star
formation history. This configuration results from the fact that the smaller
the gas fraction which is transformed into stars, the more the (ejected) ISM
contains $\alpha$-element rich SNII products.

\section{Results}
\subsection{Model constraints}
In order to convolve the galaxy models of different metallicity with the LF,
I adopt the $\langle Z\rangle$-$M_B$ relation from
\cite{ZKH94}. The mass fraction ejected is given by $M_{\rm ej}=1-f_{\rm trans}$
under the assumption that clusters basically contain gas-poor spheroidal
systems. The baryonic mass to light ratios are taken from theoretical
metallicity and age dependent SSP models (Maraston 1998)\nocite{Ma98}.

Clusters which are more rich and luminous contain a larger amount of iron.
Together with the iron mass to light ratio $M_{\rm Fe}/L_{\rm tot}$ as
introduced by \cite{Cetal91}, the slope of this relation is reproduced in
good agreement with the cluster data. The calculated iron mass to stellar
mass ratio $M_{\rm\-Fe}/M_*=0.002$ perfectly covers what is constrained in
\cite{Aetal92}. The total amount of intracluster gas, instead, is only
matched for fainter clusters like Virgo. The increase of gas ejected with
cluster luminosity is shallower than determined by observations. A steeper
slope in accordance with the data is achieved if $\beta=-1.8$, which fits
the {\em field} LF data from \cite{Lo97}. For the {\em cluster} LF, $2/3$ of
the intergalactic gas in rich clusters is predicted not to participate galaxy
formation, and must therefore have primordial origin. The [Mg/Fe] ratio in
the ejecta of the composite galaxy population is [Mg/Fe]$=-0.1$ dex, which
is a bit low but still within the range allowed by {\em ASCA} data.

\subsection{The luminosity function}
\begin{figure}[ht]
\psfig{figure=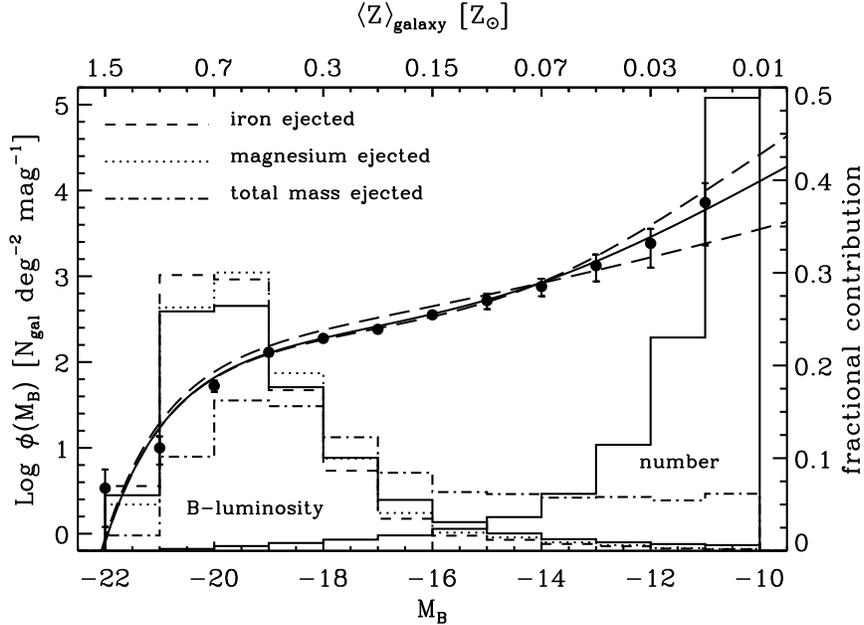,width=\linewidth}
\caption{Luminosity function (left-hand y-axis) from Trentham (1998) composite
cluster data. The fit (solid line) yields the following values for the
parameters of (1): $M^*=-20.3$, $M_{\rm t}=-14.1$, $\alpha=-1.2$,
$\beta=-0.7$. The two long-dashed lines show the fit for $\beta=-0.3$ and
$\beta=-0.9$, respectively. The fractional contributions (right-hand y-axis)
from the individual luminosity bins (respective galaxy metallicity on the
upper x-axis) to total $L_B$ and number of the galaxies are indicated by the
solid histograms. The distributions of the ejecta of gas, iron and magnesium
are shown by the histograms of different linestyles.}
\label{fig:lum}
\end{figure}
Fig.~\ref{fig:lum} shows the \cite{Tr98} LF and the respective contributions
from the galaxies of different $L_B$ to $L_{\rm tot}$ and number of the
galaxies, and to the masses of gas, iron, and magnesium in the ICM.

In spite of the huge number of dwarf galaxies, the main fraction of
$L_{\rm\-tot}$ comes from galaxies around $L^*$ ($M^*\approx -20$). Since in
the present model the production of metallicity is directly related to
galaxy mass (hence $L_B$), the distributions of iron and magnesium follow
that of $L_B$. The mass of gas delivered from the galaxies to the ICM,
instead, is much wider spread over the galaxy population. Since metal-poor
dwarf systems are less efficient in processing gas to stars, they expel a
much larger fraction of their mass than gEs. Finally, it should be
emphasized that 63 per cent of the iron in the ICM comes from SNIa, which
leads to the nearly solar Mg/Fe ratio predicted by the model. The
galaxy-cluster asymmetry is weakened, due to the significant contribution
from galaxies fainter than $L^*$.

\section{Conclusions}
For a given galaxy LF observed in clusters I analyse the fractional
contributions to the ICM enrichment from galaxies of the different
luminosity bins. The LF covers the whole relevant luminosity range from
$M_B=-22$ to $M_B=-10$, which allows to investigate the significance of
dwarf galaxies in a straightforward way. The galaxy evolution model is
semi-empirical in the sense that the relation between metallicity and
luminosity of the galaxies is taken from observational data.

It turns out that galaxies fainter than $L^*$ down to $M_B\approx -14$
provide an important contribution to the total light of the galaxy
population and to the metals found in the ICM. Dwarf galaxies below this
threshold, instead, deliver a significant fraction of the total gas mass
that is ejected by the galaxy population. Due to the influence of
intermediate ellipticals, the resulting Mg/Fe ratio in the ICM can be
reproduced consistent with {\em ASCA} measurements. SNIa are predicted to
produce 63 per cent of the ICM iron, which is a proportion comparable to the
solar neighbourhood chemistry as already claimed by \cite{Re97}. Hence, the
ICM is not an archive of SNII nucleosynthesis alone.

\section*{Acknowledgments}
I am very grateful to R.\ Bender and L.\ Greggio for the important input on
the subject. Many thanks also to A.\ Renzini, U.\ Hopp, J.\ Beuing, and C.\
Maraston for a lot of interesting discussions. In particular I would like to
thank A.\ Renzini for his crucial contribution to the nascency of this
paper. This work was supported by the "Sonderforschungsbereich 375-95 f\"ur
Astro-Teilchenphysik" of the Deutsche Forschungsgemeinschaft.


\begin{thebibliography}{}

\bibitem{}{Aetal92}{Arnaud et~al. (1992)}
Arnaud M., Rothenflug R., Boulade O., Vigroux L., Vangioni-Flam E. (1992):
A\&A {\bf 254}, 49

\bibitem{}{Cetal91}{Ciotti et~al. (1991)}
Ciotti L., D'Ercoloe A., Pellegrini S., Renzini A. (1991):
ApJ {\bf 376}, 380

\bibitem{}{Detal94}{Driver et~al. (1994)}
Driver S.~P., Phillips S., Davies J.~I., Morgan I., Disney M.~J. (1994):
Nature {\bf 268}, 393

\bibitem{}{FTH98}{Ferguson et~al. (1998)}
Ferguson H.~C., Tanvir N.~R., von Hippel T. (1998):
Nature {\bf 391}, 461

\bibitem{}{GLM97}{Gibson et~al. (1997)}
Gibson B.~K., Loewenstein M., Mushotzky R.~F. (1997):
MNRAS {\bf 290}, 623

\bibitem{}{GM97}{Gibson and Matteucci (1997)}
Gibson B.~K.,  Matteucci F. (1997):
MNRAS {\bf 291}, L8

\bibitem{}{IA97}{Ishimaru and Arimoto (1997)}
Ishimaru Y.,  Arimoto N. (1997):
PASJ {\bf 49}, 1

\bibitem{}{KC98}{Kauffmann and Charlot (1998)}
Kauffmann G.,  Charlot S. (1998):
MNRAS {\bf 294}, 705

\bibitem{}{Lo97}{Loveday (1997)}
Loveday J. (1997):
ApJ {\bf 489}, 29

\bibitem{}{Ma98}{Maraston (1998)}
Maraston C. (1998):
MNRAS in press, astro-ph/9807338

\bibitem{}{Metal97}{Mendez et~al. (1997)}
Mendez R.~H., Guerrero M.~A., Freeman K.~C., Arnaboldi M., Kudritzki R.~P.,
  Hopp U., Cappacioli M., Ford H. (1997):
ApJ {\bf 491L}, 23

\bibitem{}{MKL96}{Moore et~al. (1998)}
Moore B., Katz N., Lake G. (1998):
ApJ {\bf 457}, 455

\bibitem{}{Metal96}{Mushotzky et~al. (1996)}
Mushotzky R., Loewenstein M., Arnaud K.~A., Tamura T., Fukazawa Y., Matsushita
  K., Kikuchi K., Hatsukade I. (1996):
ApJ {\bf 466}, 686

\bibitem{}{Re97}{Renzini (1997)}
Renzini A. (1997):
ApJ {\bf 488}, 35

\bibitem{}{Retal93}{Renzini et~al. (1993)}
Renzini A., Ciotti L., D'Ercole A., Pellegrini S. (1993):
ApJ {\bf 419}, 52

\bibitem{}{Sch76}{Schechter (1976)}
Schechter P. (1976):
ApJ {\bf 203}, 297

\bibitem{}{TGB98a}{Thomas et~al. (1998a)}
Thomas D., Greggio L., Bender R. (1998a):
MNRAS {\bf 296}, 119

\bibitem{}{TGB98b}{Thomas et~al. (1998b)}
Thomas D., Greggio L., Bender R. (1998b):
MNRAS in press, astro-ph/9809261

\bibitem{}{Tr94}{Trentham (1994)}
Trentham N. (1994):
Nature {\bf 372}, 157

\bibitem{}{Tr98}{Trentham (1998)}
Trentham N. (1998):
MNRAS {\bf 294}, 193

\bibitem{}{ZKH94}{Zaritsky et~al. (1994)}
Zaritsky D., Kennicut R.~C., Jr., Huchra J.~P. (1994):
ApJ {\bf 420}, 87

\bibitem{}{Zetal97}{Zucca and et~al. (1997)}
Zucca E.,  et~al.  (1997):
A\&A {\bf 326}, 477

\end{thebibliography}

\end{document}